# Nanocavity hardening: impact of broken bonds at the negatively curved surfaces

Yu Ding,[1] Yi Chun Zhou,[2] Chang Q. Sun[1][*]

[1]School of Electronic and Electrical Engineering, Nanyang Technological University of Singapore, 639798

[2]Key laboratory of low-Dimensional Materials and Application Technologies (Xiangtan University), Ministry of Education, Hunan 411105, China

It is expected that atomic vacancies or nanometric cavities reduce the number of chemical bonds of nearby atoms and hence the strength of a voided solid. However, the hardness of a porous specimen does not always follow this simple picture of coordination counting. An introduction of a certain amount of atomic vacancies or nanocavities could, instead, enhance the mechanical strength of the porous specimen. Understanding the mechanism behind the intriguing observations remains yet a high challenge. Here we show with analytical expressions that the shortened and strengthened bonds between the under-coordinated atoms and the associated local strain and energy trapping [Sun, *Prog Solid State Chem* 35, 1-159 (2007)] in the negatively curved surface skins dominate the observed nanocavity hardening. Agreement between predictions and the experimentally observed size-dependence of mechanical strength of some nanoporous materials evidences for the proposed mechanism.

---

[*] Electronic address: ecqsun@ntu.edu.sg; zhouyc@xtu.edu.cn
URL: http://www.ntu.edu.sg/home/ecqsun/

**Introduction**

It has long been puzzling that atomic vacancies or point defects can act as pinning centers inhibiting the motion of dislocations and hence enhancing the mechanical strength of a material.[1] For examples, the hardness of FeAlN is proportional to the square root of the concentration of nitrogen vacancies.[2] The hardness of WAlC compounds increases monotonically up to a maximum at 35% C vacancies whereas the mass density decreases.[3] Fracture measurement and modeling analysis indicated that a small number of atomic defects could improve the strength of $WS_2$ nanotubes.[4] A study using atomistic simulations and analytical continuum theory[5] on the influence of the vacancy concentration on the Young's modulus and tensile strength revealed the enormous impact of an atomic defect on the strength of the nanotubes. Moreover, presence of nanometer-sized cavities also enhance the mechanical properties of solid materials.[6,7] For instance, the internal stress of an amorphous carbon film can be raised from 1 to 12 GPa by producing nanometric pores using noble gases (Ar, Kr, and Xe) bombardment during film deposition.[8,9]

Metal foams with excessive amount of discretely distributed nanocavities have formed a new class of materials, which offer a variety of applications in fields such as lightweight construction or crash energy management.[10,11] Despite the geometrical shapes of the pores,[12,13,14] the significance of the nanoporous foams is the large portion of under-coordinated atoms in the surface skins of various curvatures. The foams can be envisioned as a three-dimensional network of ultrahigh-strength nanowires or ligaments or spherical holes in the matrix. The foamed materials are expected stiffer at low temperatures and tougher at raised temperatures compared with bulk crystals. Stiffness measurement for the typical open cell Au foams of a ~30% relative density samples with different ligament sizes[15] demonstrates that the sample surface is stronger and the foams made of the smaller ligaments are even stronger.

Characterization[16] of the size-dependent mechanical properties of nanoporous Au using a combination of nanoindentation, column pillar micro compression, and molecular dynamics (MD) simulations suggested that nanoporous gold could be as strong as bulk Au, and that the ligaments in nanoporous gold approach the theoretical yield strength of bulk gold, or even harder.[15] At a relative density of 42%, porous Au manifests a sponge-like morphology of interconnecting ligaments on a length scale of ~100 nm. The material is polycrystalline with grain sizes of 10-60 nm. Microstructure characterization of residual indentation reveals a localized densification via ductile (plastic) deformation under compressive stress. A mean hardness of 145 MPa and a Young's modulus of 11.1 GPa has been derived from the analysis of the load-displacement curves. The hardness of the investigated nanoporous Au has a value some 10 times higher than the hardness predicted by the scaling laws for the open-cell foams.[17] The compacted nanocrystalline Au ligaments exhibit an average grain size of < 50 nm and hardness values ranging from 1.4 to 2.0 GPa, which are up to 4.5 times harder than the polycrystalline Au.[18] Using scaling laws for foamed materials, the yield strength of the 15 nm diameter ligaments is estimated to be 1.5 GPa, close to the theoretical strength of Au. This value agrees well with extrapolations of the yield strength in the Hall-Petch relation (HPR) at submicron scales.[19] Similarly, the strength of Al foams can be increased by 60–75% upon thermal treatment and age hardening after foaming.[20] It was also found that the hardness of the Al foam is twice as high as pure Al, and the hardness decreases with increasing temperature.[21]

On the other hand, the porous structure is thermally less stable. MD simulations[22] of the size effect on melting in solids containing nanovoids revealed four typical stages in void melting that are different from the melting of bulk materials or nanoparticles. Melting in each of the stages is governed by the interplay among different thermodynamic mechanisms arising from the changes in the interfacial free energies, the curvature of the interface, and the elastic energy induced by the density change at melting. As a result, the local melting temperatures show a strong dependence on the void size. Despite these exciting prospects, the understanding of the mechanical and thermal behavior of metal foams at the nanoscale is still very much in its infancy.[17,23]

There have been several models regarding the cavity hardening of nanovoided systems. Quantize fracture mechanics in terms of the classical continuum medium mechanics and the thermodynamic Gibbs free energy considers that a discrete number of defects arising from a few missing atoms in a nanostructure could contribute to the mechanical strength.[4, 24] Another theoretical approach considers the electronic structure around the Fermi energy.[25] Theoretical calculations suggested that the presence of two unsaturated electronic bands near the Fermi level responding oppositely to shear stress enhances the hardness of the voided systems behaving in an unusual way as the number of electrons in a unit cell changes. This finding agrees with the bond-order-length-strength (BOLS) correlation mechanism[26] indicating that a given density of states will shift to lower energy because of the broken bond depressed potential well of trapping.

According to the empirical models of foam plasticity,[18,27] the relationship between the yield strength ($\sigma$) and the relative density ($\rho_f/\rho_b$) of a foamed material follows the scaling laws,

$$\sigma_f = \sigma_b \times \begin{cases} (\rho_f/\rho_b)^{3/2} & (Gibson\ \&\ Ashby) \\ C_b(\rho_f/\rho_b)^{3/2} & (Hodge,\ et\ al) \end{cases},$$

(1)

where the subscripts $f$ and $b$ denote foam and bulk properties, respectively. The $\rho_f$ = ($V_{total}$-$V_{void}$)/$V_{total}$. Substituting the Hall-Petch relation $\sigma_b = \sigma_0 \left(1 + AK_j^{-0.5}\right)$ for the $\sigma_b$ in the modified scaling relation with a given porosity, Hodge et al[18] derived information of size dependence of ligament strength in Au foams, which follow the HPR relation with $C_b$ = 0.3 as a factor of correction. $K_j$ is the dimensionless form of solid size.

The Young's modulus of Pd and Cu foams varies with the porosity in the empirical relations,[28,29,30]

$$Y = Y_0 \times \begin{cases} 1 - 1.9p + 0.9p^2 & (\text{Wachtman \& MacKenzie}) \\ (1 - p/p_0)^n & (\text{Rice}) \end{cases}$$

(2)

with $p$ the porosity being defined as $p = V_{void}/V_{total}$. The mass density is related to the porosity in the form of $\rho_f$ = 1-$p$. The $p_0$ is the value of $p$ for which the porosity dependent properties go to zero.[31] The index $n$ and $p_0$ are adjustable parameters. A linear fit with n = 1 to the measured data of various pores has been realized using this model. The decrease in Young's modulus and flow stress with density at larger pore sizes follow exceedingly well the scaling laws attributing the observations to the existing pores that provide initiation sites for failure.

The theories given in eqs (1) and (2) have been successfully used to describe the deformation behavior of multiphase materials of larger pore sizes showing that the strength of foam materials always decreases when the porosity is increased. However, neither the effect of pore size nor the effect of bond nature of the matrix is involved in the models. Because the mechanical behavior of a surface is different from the bulk interior,[32,33,34] it would be necessary to consider the effective elastic constants of a nanofoam in terms of a three-phase structure, i.e., the bulk matrix, the voids, and the interface skins.[35] In fact, mechanical measurements of nanofoams on a submicron scale[36,37] revealed close resemblance of the nanosized ligaments in foams showing a dramatic increase in strength with decreasing ligament size.[17,19] Therefore, the effects of pore size, bond nature, temperature and in particular the role of the large portion of the under-coordinated atoms should be considered in practice. In order to apply the scaling relations to nanoporous metal foams, the yield strength should be considered as a variable of the ligament or void size. Therefore, an atomistic analysis of the effective elastic modulus of the porous systems from the perspective of bond relaxation and the associated local strain and energy trapping is necessary.

## Theory

### 1.1 Extended BOLS correlation

The core idea of the broken bond rule and the BOLS correlation mechanism[26,32] is that the broken bonds cause the remaining bonds of the under-coordinated atoms to contract spontaneously associated with bond strength gain compared with the bulk cases as standard. The shortened and strengthened bonds and the associated energy trapping dictate the unusual behavior of a mesoscopic system.

Naturally, the under-coordinated atoms surrounding atomic vacancies, point defects, nanocavities, and voids in nanofoams perform exactly the same to the under-coordinated atoms at the positively curved surfaces of nanostructures or at a flat surface despite the slight difference in the coordinating environment. The extent of mechanical enhancement or thermal stability depression is determined by the portion of the under-coordinated atoms. Therefore, we can apply directly the BOLS correlation to the negatively curved surfaces of porous structures.

## 1.2 Analytical expressions

A. Surface-to-volume ratio

Considering a sphere of $K_j$ radius with $n$ spherical cavities of $L_j$ radius lined along the $K_j$ radius, as illustrated in **Figure 1**, the entire volume $V_0$ occupied by atoms, the sum of the skins of the voids and the sphere surface, $V_i$, the porosity $p$ and mass density $\rho_f$ are calculated as,

$$\begin{cases} V_0 &= \dfrac{4\pi}{3}\left[K_j^3 - \left(n^3 + \dfrac{3}{4\pi}\right)\dfrac{4\pi}{3}L_j^3\right] & (occupied-volume) \\ V_i &= 4\pi\left[K_j^2 C_{io} + \dfrac{4\pi}{3}\left(n^3 + \dfrac{3}{4\pi}\right)L_j^2 C_{ii}\right] & (Skin-volume) \\ p &= \dfrac{4\pi}{3}\left(n^3 + 3/4\pi\right)\left(L_j/K_j\right)^3 = 1 - \rho_f & (porosity-density) \\ C_{ii} &= 2/\{1 + \exp[(12 - z_{ii})/(8z_{ii})]\} & (bond-contraction) \end{cases}$$

(3)

where $C_{ii}$ and $C_{io}$ represent the bond contraction coefficient for atoms in the inner negatively curved skins of the cavities and for atoms at the outer positively curved surface of the sphere, respectively. For the curvature dependent atomic coordination, we may extend the positive-curvature dependent coordination number to a case cover both positively (-) and negatively (+) curved surfaces:

$$z_1 = 4(1 \pm 0.75/K_j), \ z_2 = z_1+1, \text{ and } z_{i \geq 3} = 12.$$

(4)

The ratio between the volume sum of the skins and the volume entirely occupied by atoms can be derived as,

$$r_{ij}(n, L_j, K_j) = \frac{V_i}{V_0} = \frac{3}{K_j} \frac{C_{io} + 4\pi/3(n^3 + 3/4\pi)(L_j/K_j)^2 C_{ii}}{1 - 4\pi/3(n^3 + 3/4\pi)(L_j/K_j)^3} = (\gamma_{io}, \gamma_{ii})$$

$$= \frac{3}{K_j} \begin{cases} C_{io} & (Solid-sphere: L_i = 0) \\ \dfrac{C_{io} + (L_j/K_j)^2 C_{ii}}{1 - (L_j/K_j)^3} = (\gamma_{io}, \gamma_{ii})_h & (Hollow-sphere: n = 0) \\ \dfrac{3C_{io} + 4\pi(n^3 + 3/4\pi)(L_j/K_j)^2 C_{ii}}{3 - 4\pi(n^3 + 3/4\pi)(L_j/K_j)^3} = (\gamma_{io}, \gamma_{ii})_p & (Porous-sphere) \end{cases}$$

(5)

The $r_{ij} = (\gamma_{io}, \gamma_{ii})$ can be expressed in a vector form because of the coordination environment difference between the inner and the outer surfaces. The parameters $n$, $L_j$ and $K_j$ are constrained by the relation: $(2n+1)(L_j + 2) \leq K_j - 2$ because a limited number of cavities can be lined along the radius $K_j$. This expression covers situations of a solid sphere, a hollow sphere, and a sphere with uniformly distributed cavities of the same size. This relation can be extended to a solid rod, a hollow tube, and a porous nanowire as well.

With the derived surface-to-volume ratio, $r_{ij}(n, L_j, K_j)$, and the given expressions for the quantity, $q_i(z_i, d_i, E_i)$, one can readily predict the size, cavity density, and temperature dependence of a detectable quantity Q of a system with a large portion of under-coordinated atoms. The $q_i$ is the density of Q at the specific ith atomic site.

B. Thermal stability and elasticity

With the given $q_i$ relations of $T_{mi} \propto z_i E_i$, and $Y_i \propto E_i/d_i^3$ [Ref 26], we can estimate the relative change for the melting point and elastic modulus of a nanofoam to that of the bulk,

$$\frac{\Delta T_m(m, n, K_j, L_j)}{T_m(m, 0, \infty, 0)} = \sum_{i \leq 3}(r_{io}, r_{ii}) \begin{pmatrix} z_{i0b} C_{io}^{-m} - 1 \\ z_{iib} C_{ii}^{-m} - 1 \end{pmatrix}$$

$$\frac{\Delta Y(T,m,n,K_j,L_j)}{Y(0,m,0,\infty,0)} = \frac{1}{(1+\alpha T)^3} \sum_{i \leq 3}(r_{io}, r_{ii}) \left( \begin{array}{c} C_{io}^{-(3+m)}\left(1 - \frac{\int_0^T \eta_1(t)dt}{z_{iob}C_{io}^{-m}E_b(0)}\right) - 1 \\ C_{ii}^{-(3+m)}\left(1 - \frac{\int_0^T \eta_1(t)dt}{z_{iib}C_{ii}^{-m}E_b(0)}\right) - 1 \end{array} \right)$$

(6)

where m being the bond nature indicator and $E_b(0)$ the bond energy at 0 K, both are not freely adjustable parameters as both are intrinsic for a specimen. The $z_{iib} = z_{ii}/z_b$ and $z_b = 12$ is the bulk standard of atomic coordination number. $\eta_1(t)$ is the specific heat per bond, which follows Debye approximation.[32,34] The integration $\int_0^T \eta_1(t)dt$ is the internal energy of the specific bond. The calculation sums over the skin of two atomic layers.

C. Inverse Hall-Petch relationship (IHPR)

The mechanical strengthening with grain refinement in the size range of 100 nm or larger has traditionally been rationalized with the so-called T-independent HPR that can be simplified in a dimensionless form normalized by the bulk strength, $\sigma(\infty)$, measured at the same temperature and under the same conditions:

$$\sigma(K_j)/\sigma(\infty) = 1 + AK_j^{-0.5}$$

(7)

The slope A is an adjustable parameter for experimental data fitting, which represents both the intrinsic properties and the extrinsic artifacts such as defects, the pile-up of dislocations, shapes of indentation tips, strain rates, load scales and directions in the test.

As the crystal is refined from the micrometer regime into the nanometer regime, the classical HPR process invariably breaks down and the yield strength versus grain size relationship departs markedly from that seen at larger grain sizes - IHPR occurs. With further grain refinement, the yield stress peaks in many cases at a mean grain size in the order of 10 nm or so. A further decrease in grain size can cause softening of the solid, instead, and then the HPR slope turns from positive to negative at a critical size, or so-called the strongest grain size.[38] The IHPR is expressed as,[39]

$$\frac{\sigma(K_j,T)}{\sigma(\infty,T)} = \left(1 + A' \times \exp\left[\frac{T_m(K_j)}{T}\right] \times K_j^{-1/2}\right) \times \left[\frac{d(K_j)}{d}\right]^{-3} \times \left[\frac{T_m(K_j) - T}{T_m}\right]$$

(8)

where $A'$ is a prefactor and the $T_m(K_j)$ represents for the $T_m(m,n,K_j,L_j)$. The reduced bond length is given as, $d(K_j)/d = 1 + \sum_{i \leq 3}(r_{io}, r_{ii})(C_{io} - 1, \ C_{ii} - 1)*$.

Eq (8) represents that the IHPR originates from the intrinsic competition between the temperature-dependent energy-density-gain ($\propto [T_m(K_j) - T]/d^3(K_j)$) in the surface skin and the residual cohesive-energy ($\propto T_m(K_j)$) of the under-coordinated surface atoms and the extrinsic competition between activation ($\propto T_m(K_j)/T$) and prohibition ($\propto K_j^{-1/2}$) of atomic dislocations. The activation energy is proportional to the atomic cohesion which drops with solid size whereas the prohibition of atomic dislocation arises from dislocation accumulation and strain gradient work hardening which increases with the indentation depth. As the solid size is decreased, a transition from dominance of energy-density-gain to dominance of residual cohesive-energy occurs at the IHPR strongest size because of the increased portion of the under-coordinated atoms. During the transition, contributions from both processes are competitive.

**Results and discussion**

I Critical porous size

Assuming a hollow sphere of $L_j$ radius with a shell of $L_j - [L_j - (C_1+C_2)]$ thick, we have the total energy stored in the shell skin at 0 K in comparison to that in an ideal sphere without the surface effect,

$$\frac{E_{shell}}{E_{sphere}} = \frac{\int_{L_j-C_1}^{L_j} 4\pi R^2 dR C_1^{-(m+3)} + \int_{L_j-C_1-C_2}^{L_j-C_1} 4\pi R^2 dR C_2^{-(m+3)}}{\int_0^{L_j} 4\pi R^2 dR}$$

$$= \left[1 - (1 - C_1/L_j)^3\right] C_1^{-(m+3)} + \left[(1 - C_1/L_j)^3 - (1 - (C_1+C_2)/L_j)^3\right] C_2^{-(m+3)}$$

(9)

Calculations were conducted based on the given $C_i(z_i)$ and the curvature dependent $z_i$ values in eqs (3) and (4). From the results shown in Figure **2**, we can find the critical size below which the total energy stored in the shell of the hollow sphere is greater than that in the ideal bulk of the same volume without considering the temperature effects. The estimation indicates that the critical size is bond nature dependent. The critical size is 6, 8, and 11.5 for m = 1 (metal), 3(carbon, 2.56), and 5 (Si, 4.88), respectively. The elasticity of the shell is always higher than the bulk because the elasticity is proportional to the energy density. However, in plastic deformation, the hollow sphere could be stronger than the ideal bulk because of the long range effect in the indentation deformation test. On the other hand, the thermal stability of the hollow nanosphere is always lower than the solid sphere. Therefore, a hollow nanosphere should be tougher than the ideal solid sphere.

II Correlation between porosity and pore size

In Figure 3, it can be seen that the smaller the cavities the larger values of the surface to volume ratio. The properties of the porous structure are more dominated by the surface atoms for smaller cavities.

III Predictions of porosity dependence of $T_m$ and Y

Calculations of the Y and $T_m$ were conducted by using a fixed value of sphere radius $K_j$ = 600 with different $L_j$ and n values and fixed m = 1 for metals. Figure 4 shows that the $T_m$ drops when the porosity is increased; at the same porosity, the specimen with smaller pore size is less stable than the ones with larger pores. The Young's modulus increases with the porosity and the Young's modulus of the specimen with smaller pores increases faster. The predicted trends of thermal stability and strength agree well with the experiment observations for the size-dependent mechanical properties of nanoporous Au.[18,40] It is important to note that there exists porosity limit for the specimens with small pore sizes due to constrain. For the relative $T_m$ consideration, the surface-to-volume ratio should refer to the bulk volume excluding the volume of pores as given in eq (**5**); for the relative elasticity consideration, the surface-to-volume ratio should refer to the volume of the entire sphere of $K_j$ radius.

IV Plastic deformation: Inverse Hall-Petch relation (IHPR)

In dealing with the plastic deformation using IHPR, we may use the following relation to find the effective volume by excluding the pore volume in the specimen:

$$\frac{4\pi}{3}K_j'^3 = \frac{4\pi}{3}\left[K_j^3 - \left(n^3 + \frac{3}{4\pi}\right)\frac{4\pi}{3}L_j^3\right]$$

$$x = K_j'^{-0.5}$$

(10)

Figure 5(a) shows the predicted IHPR as a function of $L_j$ for $10 < K_j < 600$ specimens. Compared with the situation of single nanoparticle, the strongest size is significantly reduced for the foams. Figure 5 (b) compares the predicted IHPR of Au with experimental results. The ligament size $x(K_j^{-1/2})$ is derived from Au foams with the modified scaling relation of (1). In the figure, HPR is the classical Hall-Petch relation. IHPR 2 and IHPR 1 are the IHPR with and without involving the intrinsic competition of energy density and atomic cohesive energy as discussed for the nanoparticles. The scattered data for Au ligaments smaller than 5 nm deviates from the expected IHPR. One possibility is the surface chemical passivation effect because the higher chemical reactivity of small particles. Chemical passivation alters the bond nature of the surface bond that will enhance the strength of the bonds. A combination of the present IHPR with the scaling relation of (1) may describe the observed trends at larger porosities, and further investigation is in progress.

According to the currently developed understanding, the magnitude of $T_m - T$, or the ratio $T/T_m$, plays a key role in determining the relative strength. The $T_m$ of Al (933.5 K) is lower than that of Au (1337 K), which explains why the relative strength of Al foam to Al bulk is lower than that of Au.

IV Further evidence

The fact that the enhancement of the internal stress of a-C films by changing the sizes of nanopores through the bombardment of noble gases (Ar, Kr, and Xe)[8,9] could provide further evidence for the proposed mechanism for nanocavity hardening. The voided amorphous carbon films have an uniquely intrinsic stress (~12 GPa) which is almost one order in magnitude higher than those found in other amorphous materials such as a-Si, a-Ge, or metals (<1 GPa).[41] Using extended near-edge XAFS and XPS, Lacerda et al[8] investigated the effect of the trapping of noble gases in the a-C matrix on the internal stress of the a-C films and the energy states of the trapped gases. They found that the internal stress could be raised from 1 to 11 GPa by controlling the sizes of the pores within which noble gases are trapped. Meanwhile, they found an approximate ~1 eV lowering (smaller in magnitude) of the core level binding energy of the entrapped gases associated with 0.03-0.05 nm expansion of the atomic distance of the trapped noble gases. The measured core-level shift is of the same order as those measured for noble gases implanted in Ge,[42] Al,[43] and Cu, Ag, and Au[44,45] and Xe implanted in Pd hosts.[46] The interatomic separation of Ar (Xe) increases from 0.24 (0.29) nm to 0.29 (0.32) nm when the stress of the host a-C is increased from 1 to 11 GPa.[47]

Comparatively, an external hydrostatic pressure around 11 GPa could suppress the interplanar distance of microcrystalline graphite by ~15%,[48] gathering the core/valence electrons of carbon atoms closer together. The resistivity of a-C films decreases when the external hydrostatic pressure is increased.[49] These results are in agreement with the recent work of Umemoto et al[50] who proposed a dense, metallic, and rigid form of graphitic carbon with similar characteristics. The effect of hydrostatic pressure is very much the same as the pore-induced internal stress using noble gas sputtering and implanting.

The binding energy weakening and atomic distance expansion of the entrapped gases indicate clearly that the gas-entrapped pores expand in size and the interfacial C-C bonds contract because of the bond order loss of the interfacial C atoms, which contribute to the extraordinary mechanical strength of the entire a-C films. The pore-induced excessive stress is expected to play the same role as the external hydrostatic pressure causing densification, metallization, and strengthening of the graphite by lattice compression.

## Conclusion

It is concluded that the under-coordinated atoms in the negatively curved surfaces of atomic vacancies, point defects, nanocavities, and the syntactic foams are responsible for the strain hardening and thermal stability depression of the negatively curved systems, being the same by nature to those positively curved systems such as nanorods, nanograins and flat surfaces. Numerically, the negatively curved systems differ from the zero- or the positively-curved systems only by the fraction of the under-coordinated atoms and the coordination environment that determines the extent of BOLS induced property change. Therefore, all derivatives and conclusions for the flat surface and the positively curved surface apply to the negatively curved ones without needing any modifications though quantitative information is to be obtained both experimentally and theoretically. It is also concluded that the pores play dual roles in mechanical strength. The smaller pores act as pinning centers because of the strain and the surface trapping; the larger pores provide sites for initiating structure failure under indentation test.

**Figure captions**

Figure 1 Schematic illustration of the surface-to-volume ratio of a sphere with $4\pi n^3/3+1$ cavities and the three phase structures, i.e., voids, skins, and the matrix. Only atoms in the dark skins contribute to the property change yet atoms in the core region remain as they are in the bulk.

Figure 2 Bond nature dependence of the critical pore size below which the total energy stored in the shell of the hollow sphere is greater than the energy stored in an ideal bulk of the same size.

Figure 3 Relationship between number of cavities and porosity (a), porosity and surface to volume ratio (b) for different pore sizes of a $K_j = 600$ specimen.

Figure 4 Prediction of the porosity dependence of (a) $T_m$ and (b) Y of porous Au foams with different pore sizes of a $K_j = 600$ specimen.

Figure 5 Prediction of (a) the IHPR for nanoporous Au sphere with $10 < K_j < 600$ and different pore sizes $L_j$ and pore numbers n. (b) Comparison of the predicted IHPR of Au with measurement, Data 1 [16], Data 2 [15], Data 3 [17], and data 4 [19]. The ligament size $x(K_j^{-1/2})$ is derived from Au foams with the modified scaling relation of Ashby. HPR is the classical Hall-Petch relation. IHPR 2 and IHPR 1 are the inverse HPR with and without involving the intrinsic competition as discussed for the nanoparticles.

**Figures**

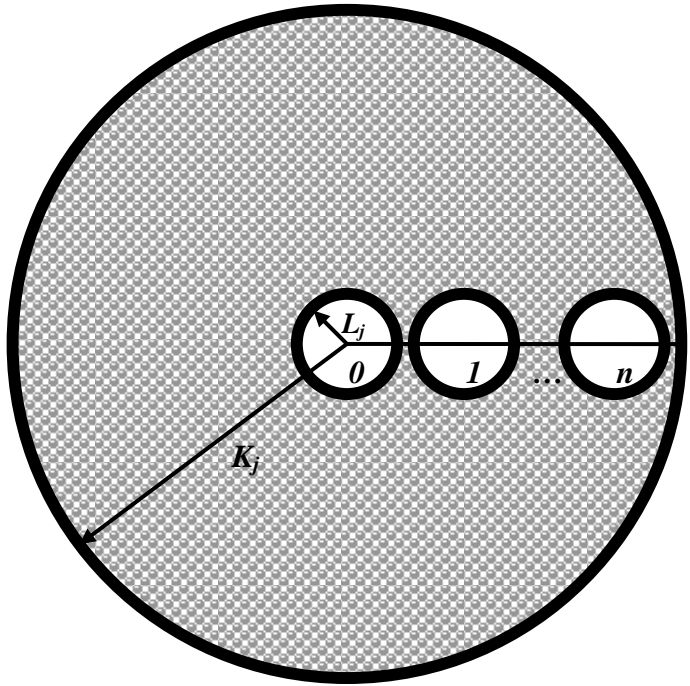

**Figure 1**

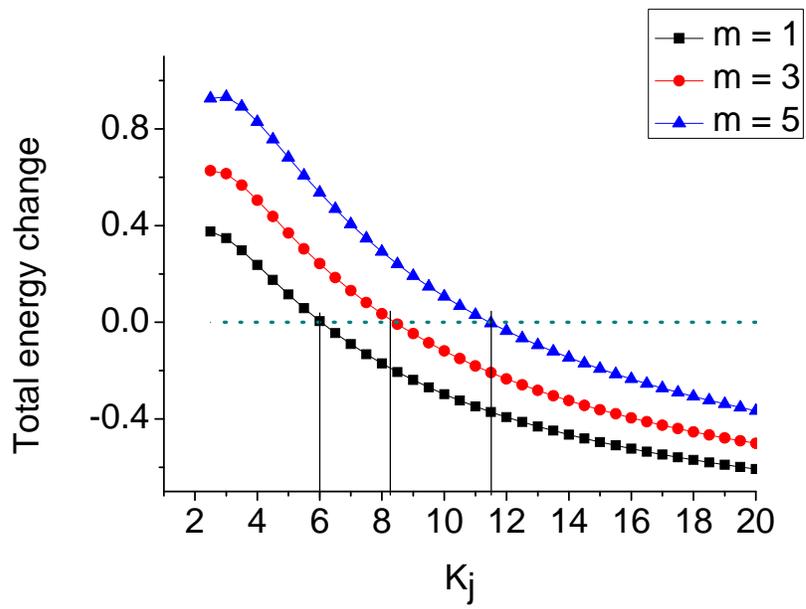

**Figure 2**

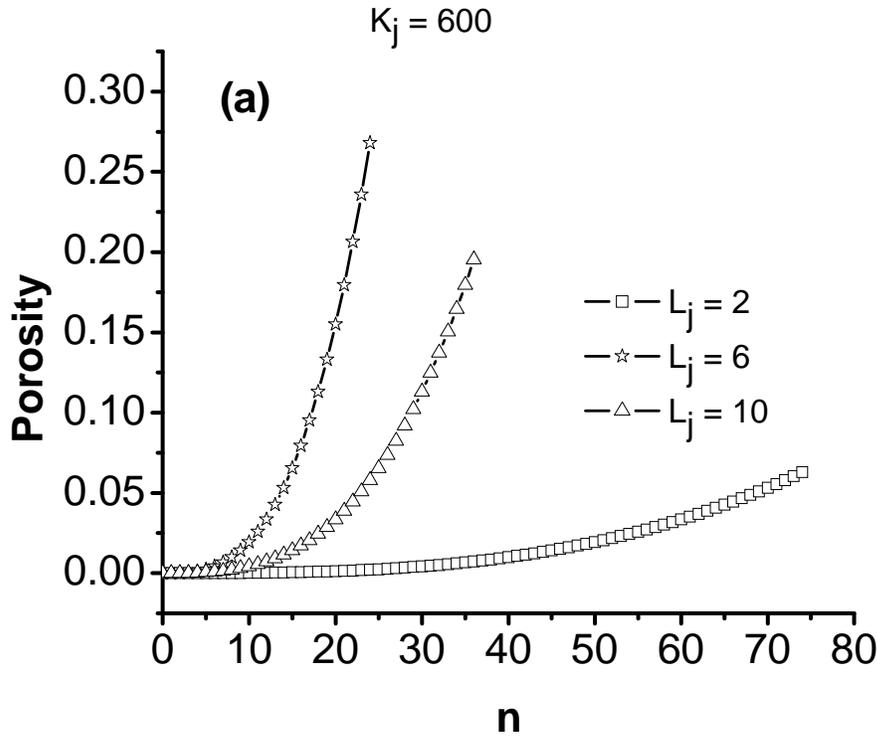

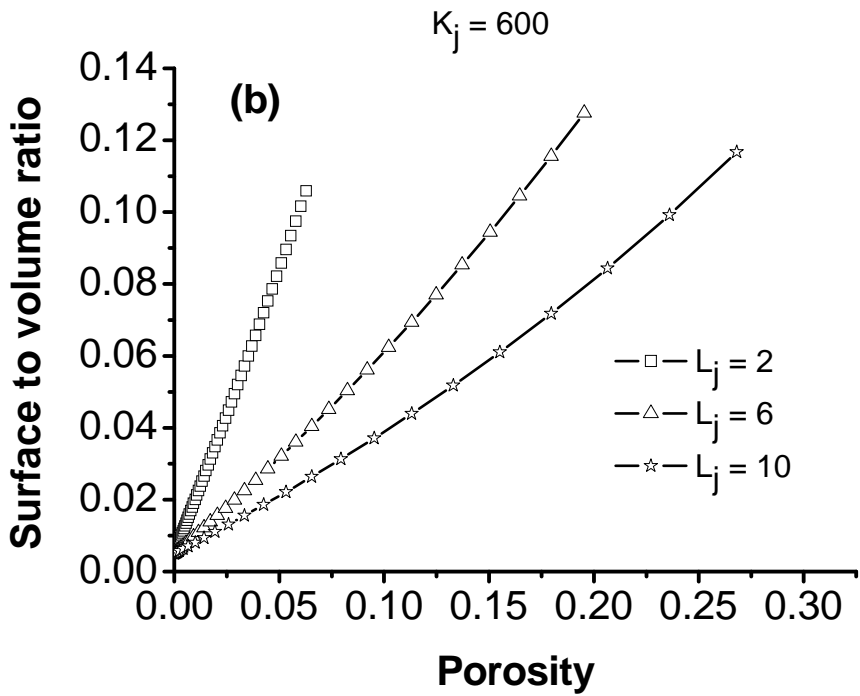

**Figure 3**

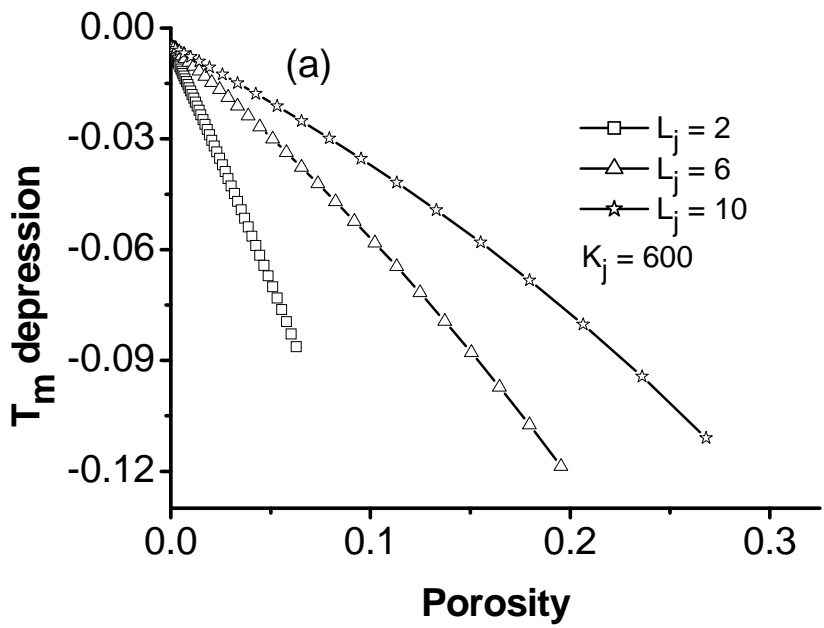

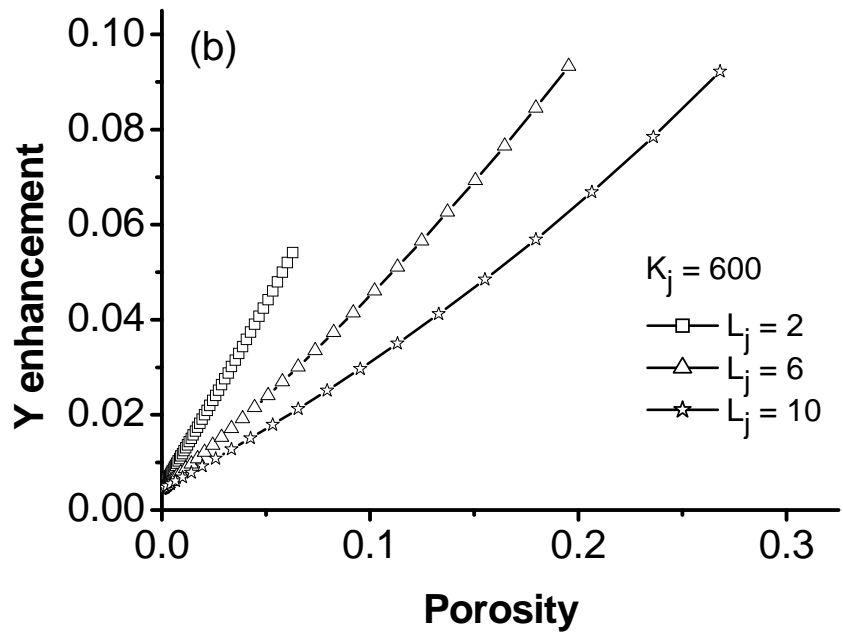

**Figure 4**

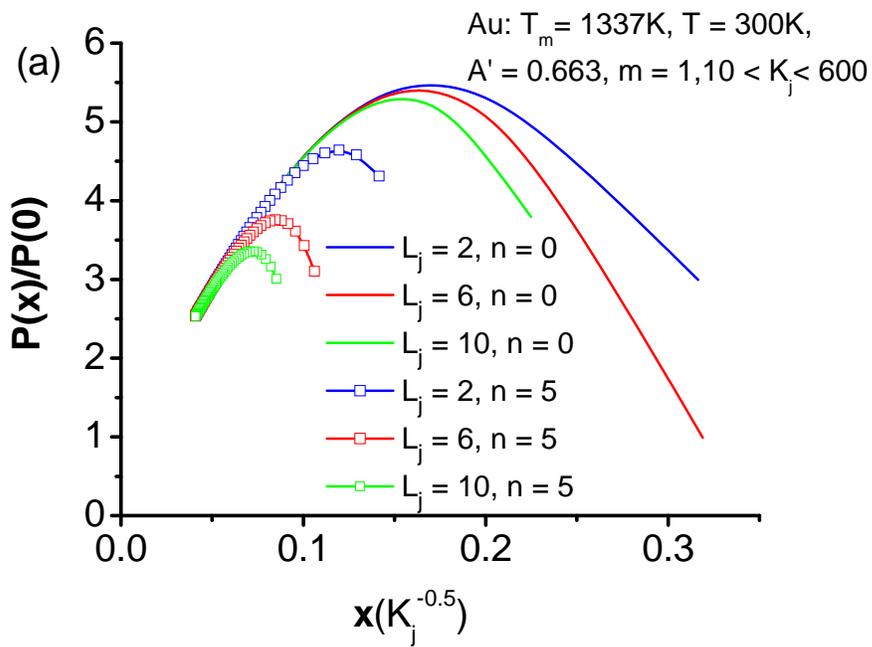
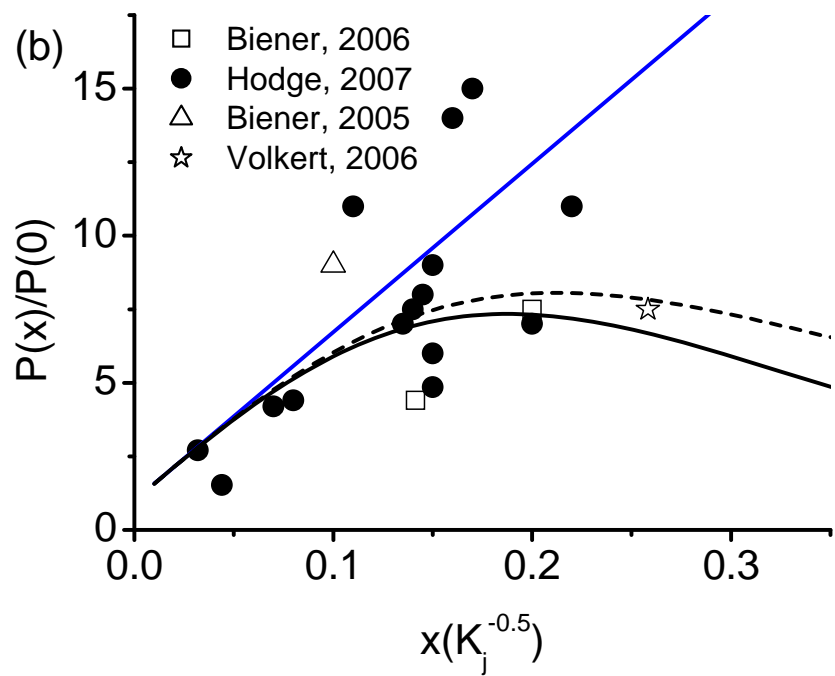

**Figure 5**